\documentclass[12pt]{article}

\usepackage[utf8]{inputenc}
\usepackage[T1]{fontenc}
\usepackage{graphicx}   
\usepackage{amsmath}    
\usepackage{amssymb}    
\usepackage{booktabs}  
\usepackage{array}
\usepackage{natbib}

\usepackage[margin=0.75in]{geometry} 
\usepackage{newtxtext, newtxmath}     
\usepackage{abstract}                

\usepackage{hyperref}   
\hypersetup{
    colorlinks=true,
    linkcolor=blue,
    filecolor=magenta,      
    urlcolor=blue,
    citecolor=blue,
}
\title{Experimental validation of an open-source low-cost single-camera 6-DOF tracking of floating-body motion in wave tanks}
\author{Herman Martens Meyer, Paul Fromont, Øystein Lande, and Atle Jensen }
\date{\today}

\begin{document}

\maketitle
\begin{abstract}


Accurately measuring motion of floating structures in experimental settings is both important and non-trivial. In this paper, we present a single-camera, 6-degree-of-freedom motion-tracking system that is both low-cost and simple to set up for an experimental campaign. The system uses fiducial markers and open-source computer vision tools to estimate the positions and orientations of multi-marker geometries and track them. We validate the accuracy and limitations on a precisely controlled linear actuator with static, regular, and irregular motion. The performance is quantified depending on both the camera-to-marker distance and direction of motion relative to the camera plane. Additionally, we present a practical use case for our system in a wave tank. Although the motion direction perpendicular to the camera plane shows the highest errors, the system achieves sub-millimeter accuracy at short and moderate camera-to-marker distances. For the irregular motion validation, the best cases reproduced the displacements with an RMSE of approximately 0.5 mm. Overall, the results indicate that low-cost single-camera fiducial-marker tracking can provide sufficiently accurate, non-intrusive motion measurements for a range of hydrodynamic laboratory experiments.

\end{abstract}

\section{Introduction}

Experimental studies of floating bodies in wave tanks rely on accurate measurements of body motion to quantify hydrodynamic interactions and validate both numerical and theoretical models. In applications such as ship model testing, wave-structure interactions, and development of floating offshore systems (floating wind turbines and wave energy converters), the six-degrees-of-freedom motion is often a key experimental output \citep{elhanafi_experimental_2017}. 

For motion measurement, several technologies are available, including commercial optical motion-capture systems such as Qualisys \citep{qualisys_system} and Vicon \citep{vicon_system}. Although these systems can provide sub-millimeter precision \citep[see, for example,][]{rhinefrank_high_2010}, they are usually expensive and require multiple cameras, which complicates setup. By contrast, sensors such as inertial measurement units (IMUs) provide a less expensive alternative, but their integration is not straightforward, and measurements can be subject to substantial noise and signal drift. These limitations motivate the development of non-intrusive, inexpensive, easy-to-deploy, and sufficiently accurate motion-tracking methods for wave-tank experiments.

In this context, computer vision-based tracking with fiducial markers offers an attractive alternative, enabling non-intrusive pose estimation with low-cost, off-the-shelf hardware and open-source software. Various types of fiducial markers are widely used in robotics \citep{kalaitzakis_fiducial_2021}. For example, \cite{Zhenglong2018} used AprilTags \citep{Olson_2011} for pose estimation on a multicopter and achieved accuracy comparable to that of a high-end motion capture system. Recently, a multi-marker 3D configuration has been shown to improve robustness over flat markers \citep{oscadal_improved_2020}, and a multi-marker approach can further enhance robustness \citep{kim_enhancing_2024}. Fiducial markers provide easily detectable geometric patterns with known dimensions, allowing the position and orientation of a rigid body to be estimated directly from images. ArUco markers \citep{GarridoJurado2014} represent one type of marker and can easily be implemented with \textit{OpenCV} \citep{opencv_library}.

Although most efforts related to fiducial markers occur in robotics, drones, and automation, some work has been done on motion tracking in marine hydrodynamics. \citet{benetazzo_accurate_2011} uses a checkerboard pattern and a single camera to obtain sub-millimeter precision in small-scale ship motion. \citet{paparella_low-cost_2019} proposes a low-cost motion-tracking system for wave-tank testing of WECs using an IMU and an ultrasound sensor, achieving millimeter accuracy, but the sensor costs around 1500 USD. \citet{ulrich_real_2023}, although not in a wave tank environment, uses an off-the-shelf camera and a WhyCode-based framework to estimate the marker positions, and reports errors of 17-35 mm and discusses the influence of camera-to-marker distance. Taken together, these studies show that, for low-cost single-camera marker-based tracking in hydrodynamic experiments, there remains limited validation of how measurement accuracy depends on camera distance, motion direction relative to the camera, and on comparing regular and irregular motion. These are factors that directly affect the usefulness of such systems in wave-tank experiments.

We present a low-cost, single-camera motion-tracking system for applications in marine hydrodynamics. We validate performance across camera-to-marker distance and motion direction relative to the camera plane for regular and irregular wave-like motion. The system is set up with a Logitech Brio 4K webcam (approx. \$130) and a multi-marker geometry of ArUco markers mounted on the tracked body. The geometry is calibrated \textit{in situ}, and the per-frame pose estimates are transformed into a common body reference frame. This work assesses whether the low-cost single-camera system can provide sufficiently accurate motion-tracking measurements for experimental hydrodynamics. The paper will discuss the practical application, validation, and limitations of the system.

The methodology is assessed through a series of validation experiments of increasing complexity. First, the measurement noise is quantified from static tests. The dynamic performance is then tested through experiments with controlled translational motion, including both sinusoidal regular and irregular motion. The irregular motion follows a JONSWAP spectrum with $T_p = 1.1$~s, representing a realistic ocean wave scenario. These measurements are repeated at several camera-to-marker distances and with motion parallel and perpendicular to the camera plane. Finally, we demonstrate a practical wave-tank application by measuring the motion of a multi-buoy wave energy converter array, showing the system's capability to simultaneously track multiple objects. Together, these experiments aim to establish the accuracy, limitations, and practical applicability of low-cost fiducial-marker tracking for floating-body motion measurements in hydrodynamic laboratories.

\section{System principle}

\begin{figure}[t!]
    \centering
    \includegraphics[width=0.8\linewidth]{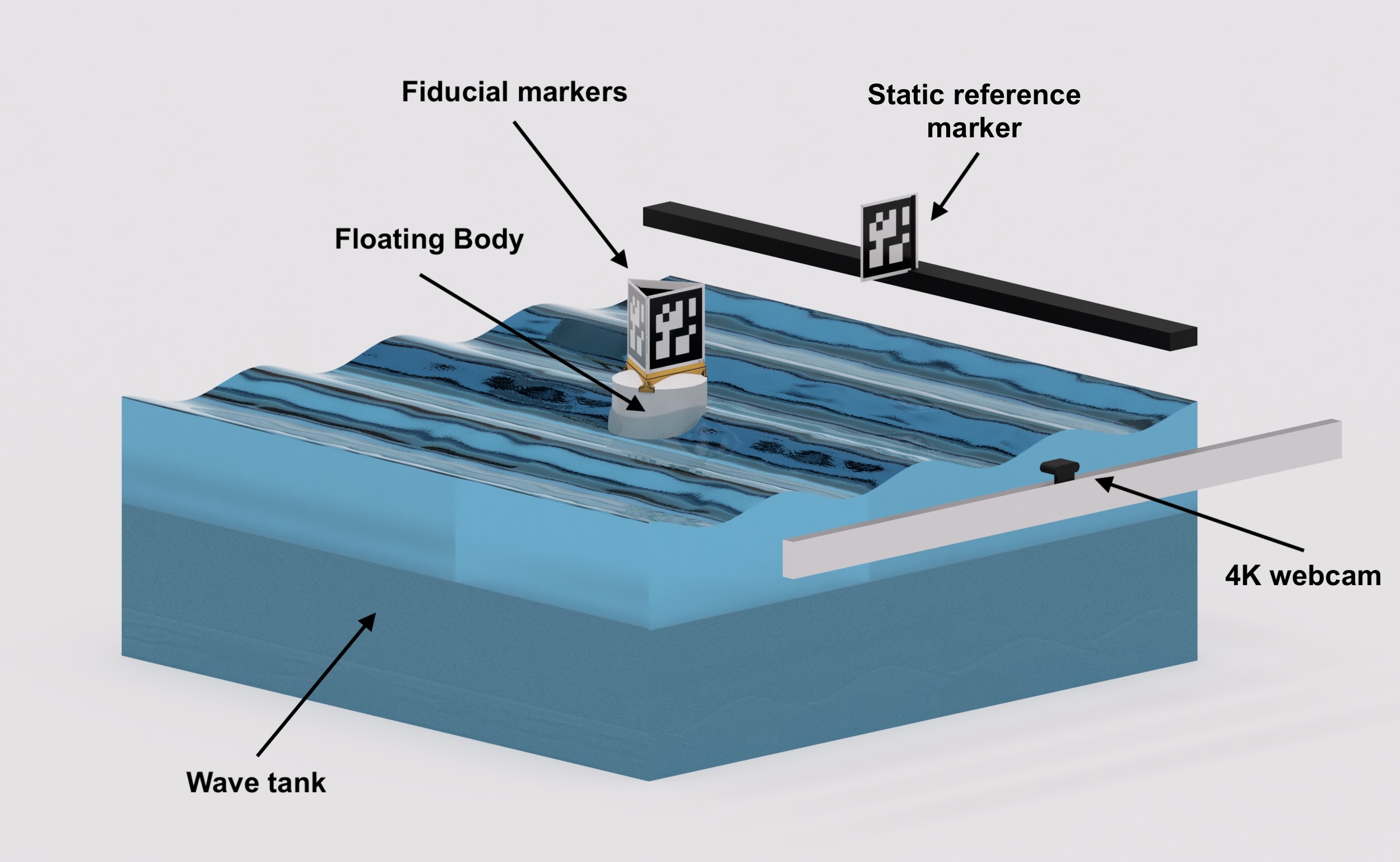}
    \caption{Schematic of the experimental setup used in the example case presented in Sec.~\ref{sec:example_WEC}, showing the camera, the floating body with multiple ArUco markers, and the static reference marker within the wave tank.}
    \label{fig:setup}
\end{figure}

\subsection{Hardware and setup}

The experimental setup required for motion tracking is described in Fig.~\ref{fig:setup}. For the validation experiments described in Secs.~\ref {sec:regular} and \ref{sec:irregular}, the moving marker is mounted on a linear actuator. The apparatus used includes a Logitech Brio 4K webcam. The shutter speed was set to 1/120 s for all validation tests. The fiducial markers are 6x6 ArUco markers \citep{GarridoJurado2014} printed at a size of $l=0.159$~m, with a static reference marker providing a stable world reference frame from which the motion of the multi-marker is referenced. The size of the markers is scaled with the camera-to-marker distance, allowing smaller markers on geometries where necessary. The multi-marker consists of three ArUco markers arranged in a triangular shape, so at least one is visible at any time. In the controlled motion experiments, the camera faced only one marker at a time, whereas in the example provided in Sec.~\ref{sec:example_WEC}, the multi-marker capability was used to its full extent.

\subsection{Workflow}
\label{sec:workflow}

\begin{figure}[ht!]
    \centering
    \includegraphics[width=0.5\linewidth]{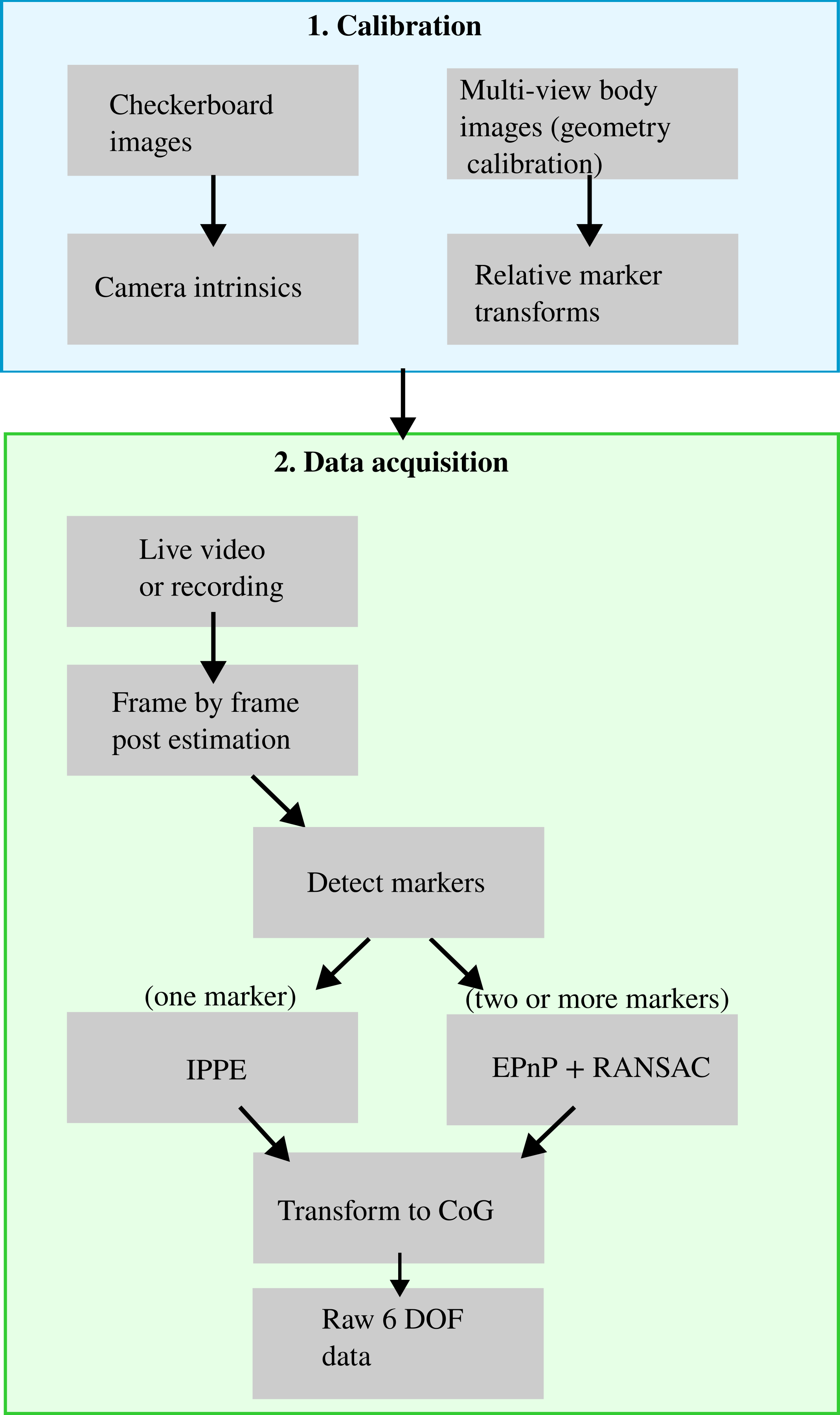}
    \caption{The workflow of the proposed tracking methodology. The process is divided into three phases: (1) calibration of camera and geometry, (2) the real-time or recorded data acquisition for per-frame pose estimation, and (3) an optional post-processing to refine the data and estimate derivatives.}
    \label{fig:workflow_version2}
\end{figure}

The complete methodology is organized into three distinct phases: (i) calibration of the camera and multi-marker geometry, (ii) frame-by-frame pose estimation from recorded images, and (iii) post-processing of the resulting trajectories. An overview of the workflow is shown in Fig.~\ref{fig:workflow_version2}, and the following sections will detail each component of this process.

\subsection{Calibration}

The calibration procedure includes two stages. First, a standard camera calibration is performed using a checkerboard pattern to estimate the camera's intrinsic parameters \citep{Zhang2000}. Subsequently, while markers can be printed and manufactured with high precision, it was deemed more practical and accurate to calibrate the exact geometric relationship between the elements of the multi-marker geometry \textit{in situ}.

\begin{figure}[htpb!]
    \centering
    \includegraphics[width=0.7\linewidth]{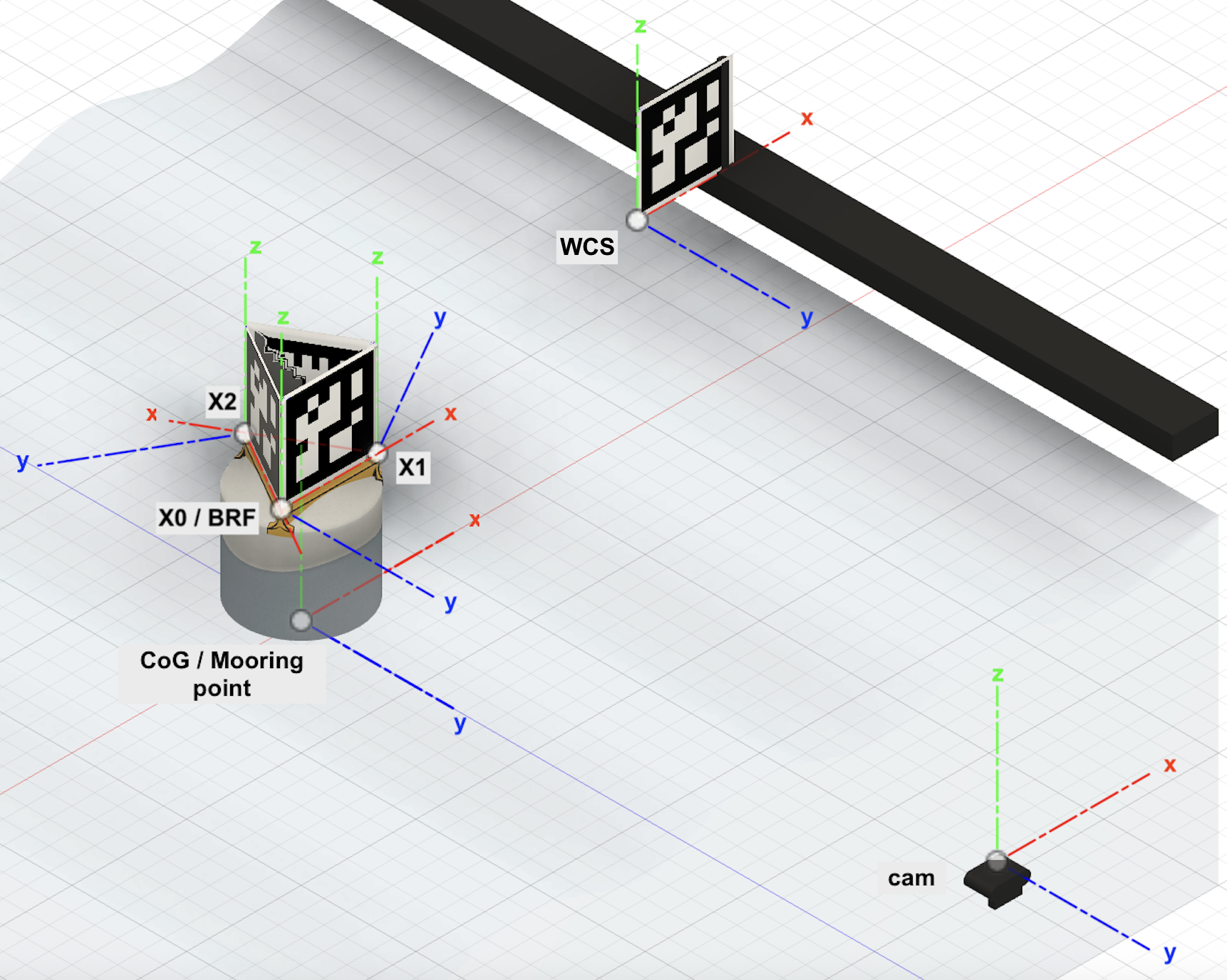}
    \caption{Diagram of the multi-marker geometry. One marker is designated as the body reference frame (BRF), $X_0$. The system calibrates the rigid transformation $\mathbf{T}_{BRF \leftarrow X_i}$ for all other markers ($X_i$) relative to the BRF.}
    \label{fig:buoy_geometry}
\end{figure}

For the triangular marker configuration exemplified in Fig.~\ref{fig:buoy_geometry}, one marker is designated as the origin of the local coordinate system, the body reference frame (BRF), denoted $X_0$. The poses of the other markers ($X_i$) on the marker geometry are then defined relative to the BRF. The multi-marker calibration includes capturing images of the marker triangle in which $X_0$ and at least one other marker, $X_i$, are simultaneously visible. For each such frame, the system estimates the pose of both markers relative to the camera, yielding the homogeneous transformation matrices $\mathbf{T}_{cam \leftarrow X_0}$ and $\mathbf{T}_{cam \leftarrow X_i}$. The constant relative transformation from marker $X_i$ to $X_0$ is then calculated by composing the transformations:
\begin{equation}
    \mathbf{T}_{BRF \leftarrow X_i} = \mathbf{T}_{X_0 \leftarrow X_i} = (\mathbf{T}_{cam \leftarrow X_0})^{-1} \cdot \mathbf{T}_{cam \leftarrow X_i}
    \label{eq:relative_transform}
\end{equation}
This procedure is repeated from multiple perspectives, and the resulting relative translations and rotations are averaged to estimate the marker geometry \citep{Markley2007}. The resulting set of transformations $\{\mathbf{T}_{BRF \leftarrow X_i}\}$ for all $i$ defines the calibrated geometry of the rigid body. The calibration step can be easily extended to different geometries with varying numbers of markers.


\subsection{Multi-marker rigid-body pose estimation}
\label{sec:rigid_body_pose}

The per-frame processing of the acquired images consists of the following steps. First, the markers are detected, and the four corners are refined to sub-pixel accuracy. Thereafter, the pose of each visible marker is estimated using Infinitesimal Plane-Based Pose Estimation (IPPE) \citep{collins_infinitesimal_2014}. Markers with poor geometry and excessive reprojection error were rejected. The corners of all accepted markers are transformed into the common body reference frame, and a rigid body pose is estimated. If two or more markers are visible simultaneously (8 corners or more), pose estimation was performed using EPnP \citep{lepetit_epnp_2009} with RANSAC-based outlier detection \citep{fischler_random_1981} using OpenCV’s \texttt{solvePnPRansac} functionality. The solution was further refined using iterative PnP and rejected when its mean reprojection error exceeded 10 pixels.

The system can run in \textit{live} mode, where frames are processed on the while recording, or it can process pre-recorded images. The \textit{live} mode may experience lower FPS processing depending on the computer. 

\section{Experimental Validation}\label{sec:validation}

To quantify the performance of the proposed methodology, a series of validation experiments was conducted, progressing from static precision assessment to dynamic tracking in a more realistic wave environment. Additionally, we present an example use case of the system for a hydrodynamic analysis of a wave energy converter. To enable a comparison for markers of different sizes and different camera-to-marker distances, we will refer to the non-dimensional camera-to-marker distance $d/l$, where $d$ refers to the camera distance and $l$ is the marker width. These values are camera-specific, and we also present the average pixel width of each marker, denoted by $\mathcal{W}$. We tested $\mathcal{W} = 360, 180,120$ pixels per marker, which corresponds to $d= 1,2,3$~m and $d/l = 6.3, 12.6, 18.9$.

\subsection{Static test}

To quantify the tracking system's measurement noise, the first segment of each validation run was analyzed. For all regular and irregular motion experiments, the first five seconds of the time series were recorded with the markers at rest. We calculated the standard deviation of the static periods across all degrees of freedom and camera-to-marker distances to quantify the static noise level.

The resulting noise levels are presented in Table \ref{tab:static_noise}. A clear trend is that translational noise is lowest for motion in the camera plane ($x, z$) and substantially higher in the transverse plane ($y$), as expected for a single-camera system. At $d/l=6.3$, in-plane noise standard deviation is approximately $0.2$~mm, increasing to around $0.3$~mm at $d/l=18.9$. In contrast, the noise in the $y$-direction increases from $1.2$~mm to $2.1$~mm over the same distance range. The rotational noise shows the same qualitative behavior, where the in-plane motion exhibits the lowest noise levels. These measurements will serve as a benchmark for the precision we can expect from the dynamic measurements.

\begin{table}[b!]
    \centering
    \caption{Static measurement noise as a function of camera distance.
    Values represent the median standard deviation across experimental runs,
    calculated from the initial static period of up to 5~s. The out-of-image-plane motion is in the $y$-direction.}
    \label{tab:static_noise}
    \begin{tabular}{@{}lccc@{}}
        \toprule
        DOF &
        $d/l=6.3$ &
        $d/l=12.6$ &
        $d/l=13.9$ \\
        \midrule
        $X$ (mm)     & 0.193 & 0.216 & 0.317 \\
        $Y$ (mm)     & 1.242 & 1.651 & 2.095 \\
        $Z$ (mm)     & 0.213 & 0.239 & 0.323 \\
        \addlinespace
        $\theta_x$ (deg) & 0.229 & 0.505 & 0.283 \\
        $\theta_y$ (deg) & 0.044 & 0.094 & 0.061 \\
        $\theta_z$ (deg) & 0.279 & 1.619 & 0.222 \\
        \bottomrule
    \end{tabular}
\end{table}





\subsection{Forced oscillation test}

Moving from a still system and noise measurement, we wanted to validate the motion-tracking ability in actual, relevant motion. The validation consisted of both regular and irregular wave motion. The Aruco markers were subjected to translational motion using a precisely controlled linear actuator. To ensure that the full capabilities of the motion-tracking system were evaluated, all motion validation tests were conducted with the camera positioned such that we measured motion parallel and perpendicular to the camera plane. Furthermore, measurements were performed at camera-to-marker distances of 1, 2, and 3 m, corresponding to $d/l = 6.3, 12.6, 18.9$. All tests were repeated three times, and the regular motion was performed over 20 periods, and the JONSWAP spectrum was 6 minutes long with a peak period $T_p = 1.1$ s.

\subsubsection{Regular motion} \label{sec:regular}

The regular-motion experiments were used to assess how accurately the tracking system reconstructs the actuator's sinusoidal motion as a function of camera-to-marker distance, frequency, and direction of motion. The experimental matrix is summarised in Tab.~\ref{tab:regular_motion_tests}. For each case, the measured marker displacement was compared with the actuator motion to assess amplitude agreement.

\begin{table}[b]
\centering
\caption{Regular motion experimental matrix used for validation of the ArUco tracking system.}
\label{tab:regular_motion_tests}
\begin{tabular}{cccc}
\hline
Amplitude, $a$ [cm] & Frequencies, $f$ [Hz] & Distances, $d/l$ & Motion directions \\
\hline
5  & 0.5, 1.0, 1.5 & 6.3, 12.6, 18.9 & $x$, $y$ \\
10 & 0.5, 1.0      & 6.3, 12.6, 18.9 & $x$, $y$ \\
15 & 0.5           & 6.3, 12.6, 18.9 & $x$, $y$ \\
\hline
\end{tabular}
\end{table}

\begin{figure}[ht]
    \centering
    \includegraphics[width=0.7\linewidth]{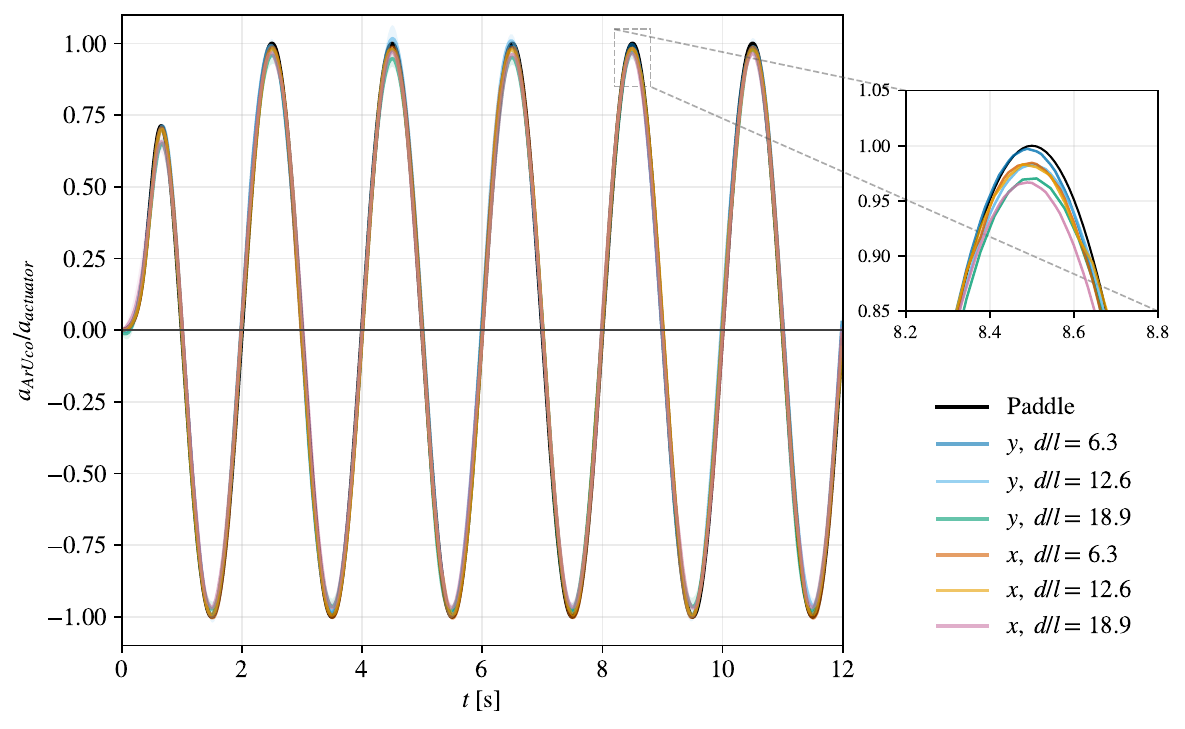}
    \caption{Comparison of the actuator motion at $a=15$ cm and $f=0.5$ Hz with measured motion the the six different measurement positions. Here, $x$ denotes motion parallel to the camera plane, and $y$ is motion perpendicular to the camera plane.}
    \label{fig:figure_regular_f0p5_a15_comparing_motion_allCases}
\end{figure}

\begin{figure}[ht]
    \centering
    \includegraphics[width=\linewidth]{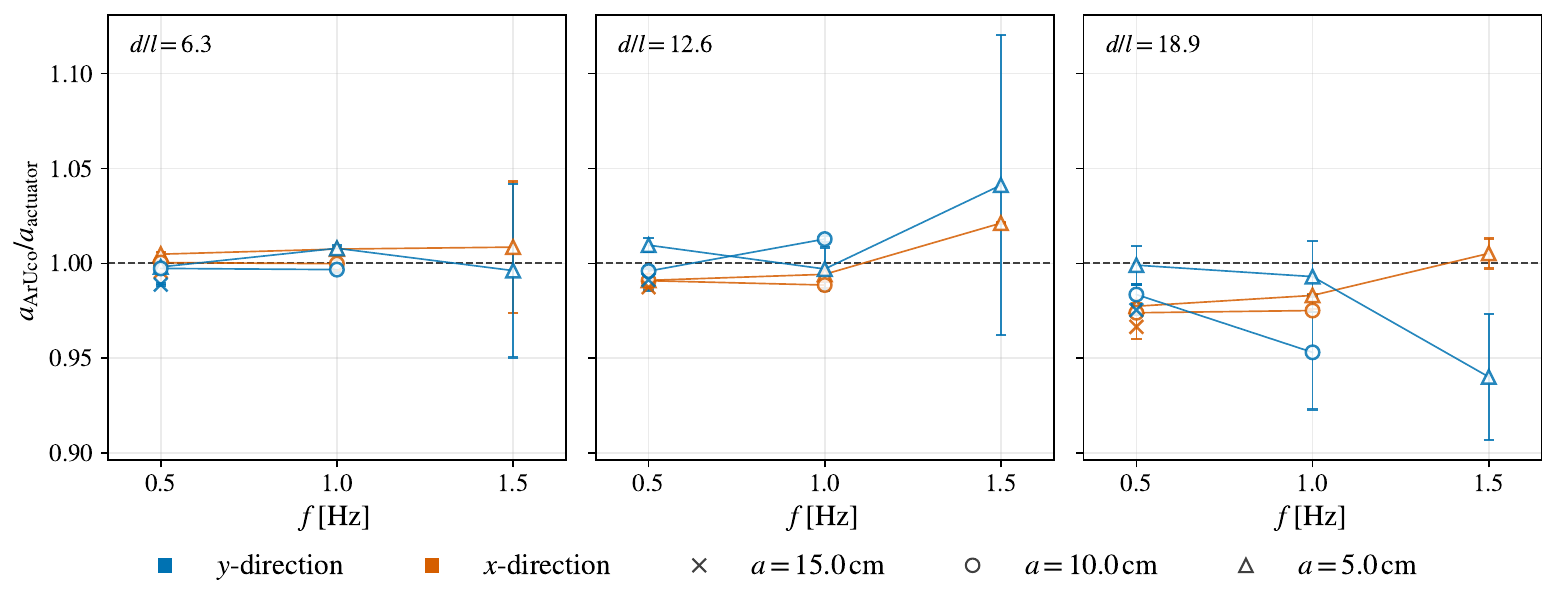}
    \caption{Normalized amplitude $a_{ArUco}/a_{actuator}$ as a function of the actuator input frequency for the three non-dimensionalized recording distances, $d/l$ = 6.3, 12.6 and 18.9. Here, $x$ denotes motion parallel to the camera plane, and $y$ is motion perpendicular to the camera plane.}
    \label{fig:figure_normalised_amplitude_vs_frequency_by_distance}
\end{figure}

A representative example is shown in Fig.~\ref{fig:figure_regular_f0p5_a15_comparing_motion_allCases} for the case $a=15$~cm and $f=0.5$~Hz. Both the period and the amplitude of the marker are accurately captured. From the inset, we see a snapshot of the observed trend: the measured amplitude is underestimated when the distance between the camera and the marker is greatest. This behaviour is likely the cumulative result of several factors. At the largest $d/l$ tested, the marker occupies only 120 pixels (compared to 360 at the closest), reducing the effective spatial resolution and increasing the physical distance per pixel. As a result, uncertainties associated with sub-pixel corner localization, edge detection, lens distortion correction, and pose estimation become more significant. When comparing all regular validation cases in Fig.~\ref{fig:figure_normalised_amplitude_vs_frequency_by_distance}, the trend of the amplitudes being underestimated at $d/l=18.9$ m is also visible. 

In general, as expected, the motion tracking is most accurate at shorter distances from the camera to the marker, and for $d/l=6.3$, we measure sub-millimeter errors in the amplitudes, which we also see in two of the cases at $d/l=12.6$. As briefly mentioned, this accuracy deteriorates as $d/l$ increases, and at $d/l=18.9$ we see the largest errors: a maximum error of around 4.8 mm for $a=15$~cm, which is related to the marker occupying a smaller area of the frame, making the displacement in pixels smaller and harder to capture.

The tracking of out-of-image-plane motion is surprisingly precise. For motion parallel to the image plane, a physical displacement produces a direct displacement of the marker in image coordinates, meaning a 5 cm motion corresponds to tens of pixels of image displacement. Conversely, motion towards or away from the camera is reflected in a change in the apparent marker size. For the values of $\mathcal{W}$ considered here, the marker size will change by only a few pixels in the out-of-image-plane motion experiments.

Another observation from the regular validation experiments is that the tracking presents more errors at the highest tested frequency. The same displacement at a higher frequency results in a faster-moving marker, and since we were restricted to 30 fps and kept the camera shutter speed at 1/120 s for all experiments, this led to blurrier images, poorer corner detection, and less precise tracking. This highlights a limitation of the current setup, while also inviting more precise measurements with a more sophisticated camera. Better lighting would also have enabled the use of a faster shutter speed, producing clearer images.

We note that at $d/l=12.6$ and $f=1.5$ Hz, for the $z$-motion case, we observed image blurring possibly related to vibrations in the tank where the camera was mounted. This explains the large error bars in this case but also underscores the need for a stable camera position. We believe this also explains the higher noise levels measured in this case, as shown in Tab.~\ref{tab:static_noise}.

\subsubsection{Irregular motion} \label{sec:irregular}

To validate irregular motion, we programmed the actuator to follow a JONSWAP spectrum (with $T_p = 1.1$~s) for 6 minutes. These spectra were repeated three times for each of the six measurement locations. The purpose was to test the motion-tracking capabilities in more realistic wave-motion conditions and to assess the ability to capture both spectral distributions and individual crests.

\begin{figure}[ht]
    \centering
    \includegraphics[width=\linewidth]{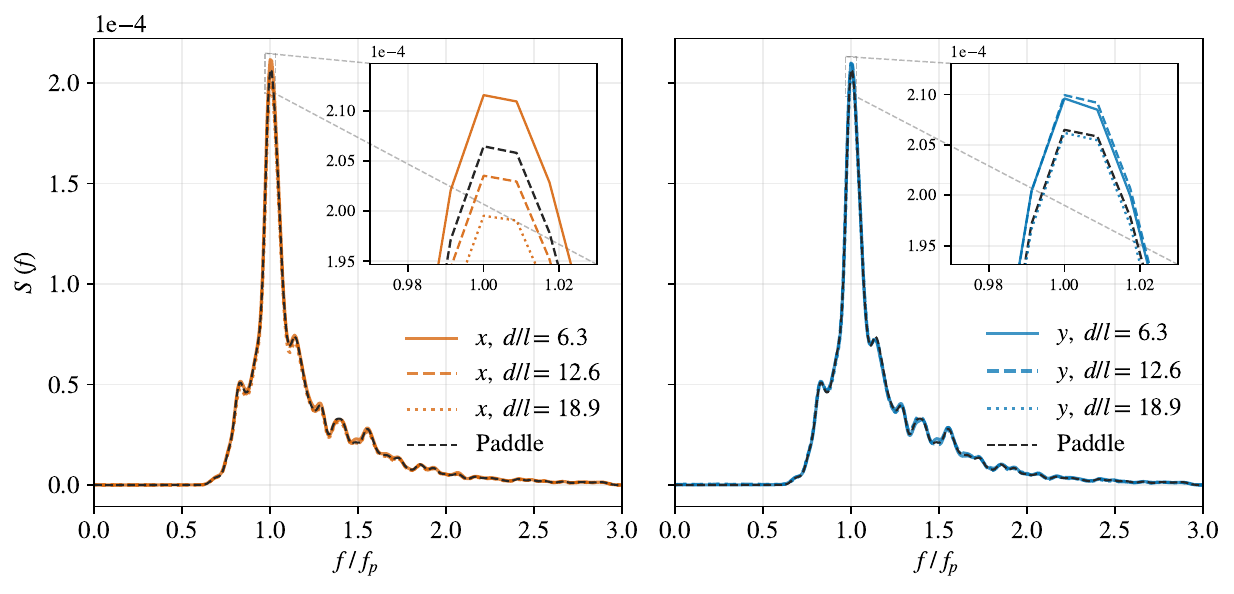}
    \caption{Comparison of the measured spectra from both the actuator signal and all experimental measurement locations. Here, $x$ denotes motion parallel to the camera plane, and $z$ is motion perpendicular to the camera plane.}
    \label{fig:figure_jonswap_compare_runs}
\end{figure}

Comparing the spectrum of the actuator motion with the experiments, we see a very accurate fit in Fig.~\ref{fig:figure_jonswap_compare_runs} for all tested cases. Zooming in on the spectral peak, it can be argued that the peak is more accurately described in the case of transversal motion, contrary to what we see in the regular motion. However, this difference is below the data scatter. In the discussion of regular waves, where we saw that the highest frequency was poorly measured, we are now seeing very accurate detection of the highest-frequency signals in the irregular validation experiments. Where we used a fixed amplitude in the regular experiments, the highest-frequency components in the JONSWAP spectrum will be at lower amplitudes, and the problems we observed with high velocities, shutter speed, and framerate are likely not as severe here.

Spectral agreement alone does not guarantee that the time-resolved motion is accurately reconstructed. We therefore also compared the measured and reference time series directly. When the measured time series is correlated with the actuator motion, the average across all runs is 0.99, except for out-of-image motion at the two largest distances, where it is 0.98. The lag between the measured motion and the actuator motion is below the capturing frequency.

\begin{figure}[ht]
    \centering
    \includegraphics[width=\linewidth]{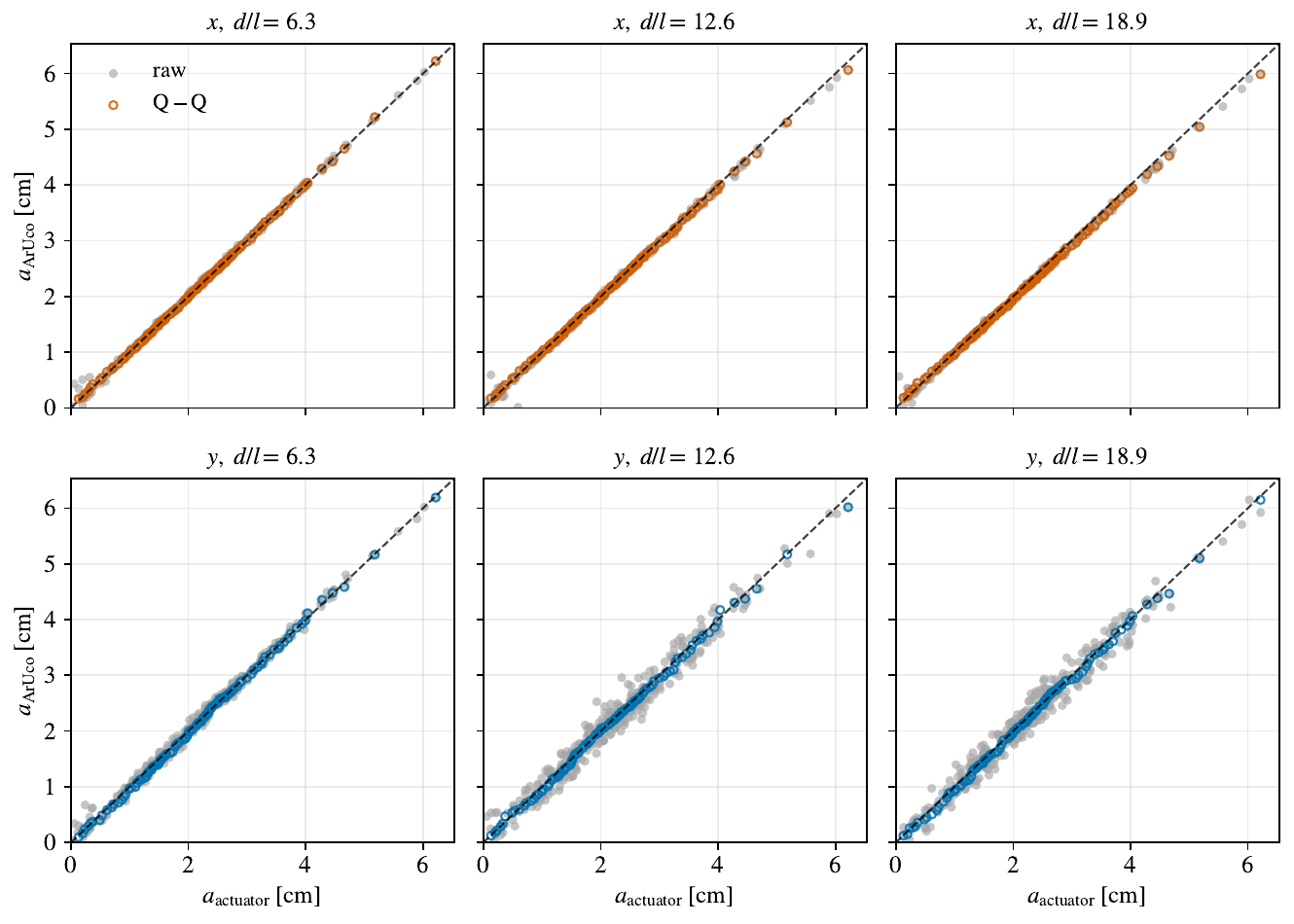}
    \caption{The gray cloud represents individual quantile pairings from all realizations, while the coloured symbols denote the ensemble-averaged quantiles for each measurement location. The dashed line indicates perfect agreement. The direction of motion is indicated by $x$ (in-plane) and $y$ (out-of-image-plane).}
    \label{fig:figure_jonswap_crest_qq_all}
\end{figure}

Furthermore, we compared the crest and trough statistics of the actuator and ArUco signals. For each irregular case, the mean was removed, and a weak high-pass filter (0.03 Hz) was applied to reduce residual drift before all extrema were extracted between zero crossings. Figure~\ref{fig:figure_jonswap_crest_qq_all} shows both the scatter plot of the time-matched extrema pairs and a quantile-quantile comparison of the resulting extrema distributions. The grey scatter points display individual actuator extrema paired with their nearest ArUco extrema in time and therefore indicate event-by-event agreement. The coloured Q-Q points compare equal percentiles of the two extrema distributions, and therefore indicate whether the statistical distribution of displacements is preserved. Both the raw scatter and the Q-Q points lie close to the one-to-one line, indicating that the tracking system captures not only the spectral content of the irregular motion but also the individual peaks. Small differences from the one-to-one line are mainly visible in the largest crests, and the scatter of the raw data is largest for the $z$-plane motion, as we would expect from both theory and from noise levels in the static test.

Looking at the RMSE of the QQ-data and the raw data, we are seeing a sub-millimetre precision in all cases, while for the raw data, RMSE is below one millimetre for all cases of in-plane motion and for out-of-image-plane motion at $d/l=6.3$ m, while the clear scatter we see in the two remaining cases has a raw RMSE of approximately 1.7 mm. Raw pair RMSE of $0.5 - 1.7$~mm. In the $z$-motion, we observe higher noise, so the accuracy is not better than 1-2 mm, depending on the camera-to-marker distance.

\subsection{Use case example - motion of wave energy converter} \label{sec:example_WEC}

\begin{figure}[ht]
    \centering
    \includegraphics[width=0.9\linewidth]{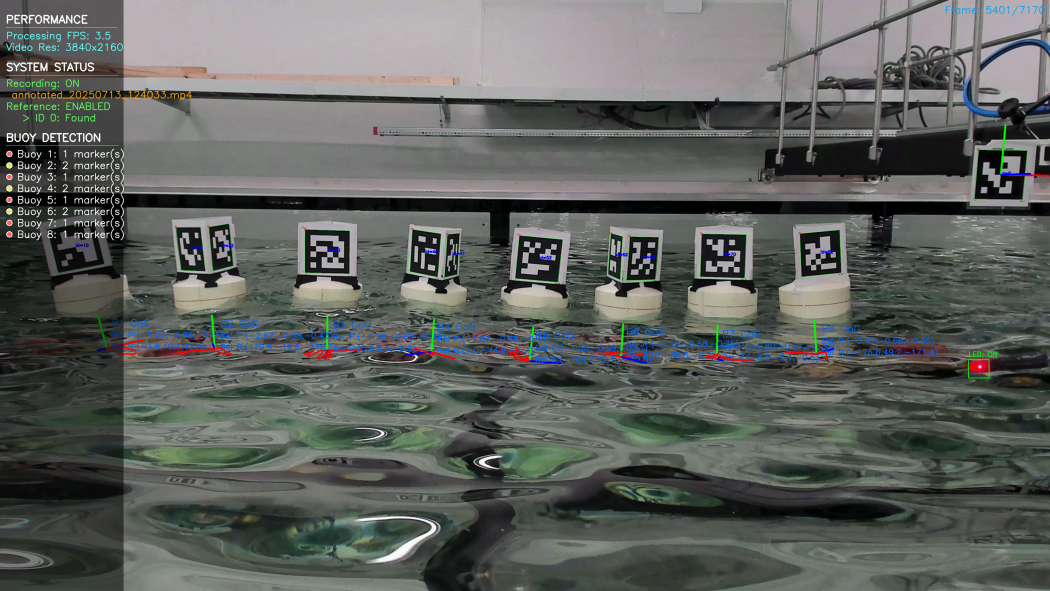}
    \caption{A snapshot of the processing of buoy motion in the WEC with the motion tracking system. The centre of gravity (CoG) is offset from the multi-marker to the CoG of the buoys.}
    \label{fig:figure_wec_case15_annotated_frame}
\end{figure}

In addition to the validation results, we present an example of how the motion-tracking system can be used in a practical setting. We present a relevant example from the ArUco tracking system use case in a wave energy converter with a buoy array aligned with wave propagation. This system is described in \cite{meyer2026experimentalinvestigationmultibuoycooperative}, and with ArUco markers on each of the eight buoys in the array, we can capture their motion responses to incoming waves. Figure \ref{fig:figure_wec_case15_annotated_frame} shows what this setup can look like for a processed frame with the motion capture system. This example highlights one of the main strengths of this system. With the ArUco markers, multiple bodies can be tracked simultaneously with their unique tags, a feature which is not necessarily straightforward with traditional IR-based tracking systems. 

\begin{figure}[ht]
    \centering
    \includegraphics[width=\linewidth]{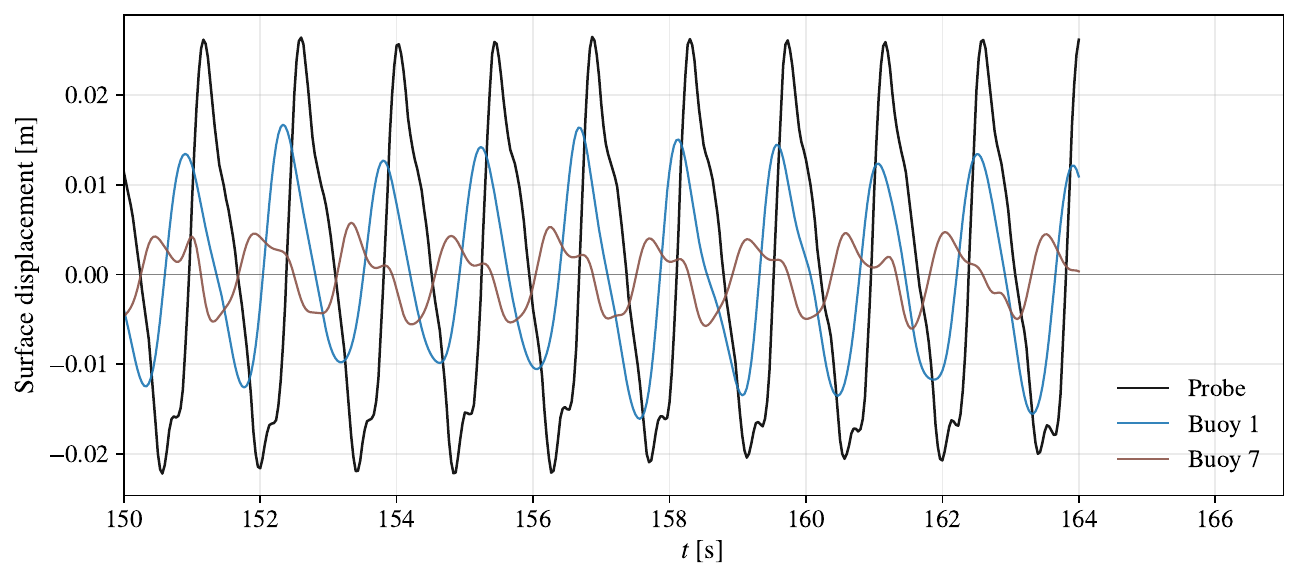}
    \caption{Comparison of the incoming wave measured with the wave probe and the heave motion of buoys 1 and 7 for a sample of the obtained time series.}
    \label{fig:figure_wec_buoy1_y_vs_Gauge_1}
\end{figure}

This example demonstrates how the proposed tracking system can be deployed in a realistic laboratory setting. As an illustration, Fig.~\ref{fig:figure_wec_buoy1_y_vs_Gauge_1} compares the incident wave elevation measured by a wave probe with the heave motion of the first and seventh buoys in the array. The example shows that the tracking system can provide time-resolved buoy-motion measurements directly comparable with wave-probe and power-take-off data and can therefore form part of a broader hydrodynamic analysis of multi-body wave energy systems.
    

\section{Conclusions}

We have presented and validated a low-cost, single-camera workflow for six-degree-of-freedom motion tracking of floating bodies in wave tank experiments using ArUco markers and open-source computer vision tools. The system combines single-camera pose estimation with an \textit{in situ} calibrated multi-marker geometry. The system was assessed using static tests and controlled actuator-motion experiments, and demonstrated through a wave-energy-converter experiment example.

The noise level, determined using static tests, is around 0.2-0.3 millimeters for translational motion in the camera plane, while the noise in the depth direction is higher, as expected for a single-camera system. The noise increases with camera-to-marker distance. In the regular motion validation, the actuator motion was reconstructed accurately across the tested amplitudes and frequencies, with the best performance observed at the shortest camera distances. The largest errors were observed at the largest tested distance at the highest marker velocities, indicating that marker image size, motion blur, frame rate, and shutter speed are important practical limitations of the present setup. 

Although out-of-image-plane motion measurements were less accurate than in-plane measurements, they remained within the millimeter range in most cases. As noted, this behavior is expected, as in-plane motion corresponds to direct image displacement, and out-of-image-plane motion is captured from small changes in the apparent marker size. Nevertheless, the system recovers three-dimensional rigid-body motion with high accuracy for the hydrodynamic applications investigated. For applications requiring higher depth accuracy, the proposed methodology could be extended to a stereo-camera configuration while preserving the same marker-based workflow.

The validation of irregular motion was particularly relevant for the intended hydrodynamic application of the method. For irregular actuator motion, the tracking system reproduced the measured spectra well across all frequencies, and the measured time series showed cross-correlation with the references above 0.98 in all tested cases. The comparison of crest and trough amplitudes showed that the magnitudes and distributions of individual extrema were measured accurately, with raw-pair RMSE of $0.5-1.7$~mm, depending on the camera-to-marker distance and motion direction.

The post-processing framework had only a small effect on the validation measurements, indicating that the raw motion tracking is robust under the tested conditions. Its main value is therefore expected to be greater in experiments with stronger noise and missing data due to marker occlusion.

Finally, the demonstration of a multi-buoy wave energy converter showed how the tracking workflow can be deployed in a realistic wave tank setting to obtain synchronized motion measurements of multiple floating bodies. While a full hydrodynamic analysis of that system is beyond the scope of the present paper, the example illustrates the method's practical use in experimental studies of multiple floating structures simultaneously.

The validation shows that low-cost motion tracking of ArUco markers with a single camera can achieve sub-millimeter accuracy across a wide range of hydrodynamic applications, provided that limitations related to camera-to-marker distance, multi-marker geometry, and image quality are accounted for. The method is non-intrusive and simple to deploy, and therefore offers an alternative to more expensive motion-capture systems in hydrodynamic laboratory experiments.



\section{Acknowledgements}

We are grateful for the contributions of Théo Krzywkowski and Olav Gundersen. Øystein Lande has received funding from the Research council of Norway and Ocean Oasis AS. We used LLM-powered tools to improve the language of the manuscript. All scientific content is the work of the authors, who reviewed and approved all LLM-assisted language edits.

\newpage
\bibliographystyle{plainnat}
 \bibliography{bibliography.bib}

\appendix

\end{document}